\documentclass[11pt]{article}
\usepackage{epsfig}
\usepackage[usenames]{color}
\usepackage{amsmath}
\usepackage{amsfonts}
\usepackage{amssymb}
\usepackage{mathrsfs}
\renewcommand{\theequation}{\mbox{\arabic{section}.\arabic{equation}}}

\newtheorem{proposition}{Proposition}[section]
\newcommand{\bpr}{\begin{proposition}}
\newcommand{\epr}{\end{proposition}}

\newcounter{Roman}
\addtocounter{Roman}{1}

\newcommand{\beq}{\begin{equation}}
\newcommand{\eeq}{\end{equation}}
\newcommand{\bea}{\begin{eqnarray}}
\newcommand{\eea}{\end{eqnarray}}
\newcommand{\bml}{\begin{multline}}
\newcommand{\bal}{\begin{align}}
\newcounter{saveeqn}

\newcommand{\D}{\displaystyle}

\newcommand{\ssc}{\scriptscriptstyle}

\newcommand{\hl}{\hat{l}}

\newcommand{\rs}{{\rm s}}
\newcommand{\rw}{{\rm w}}

\newcommand{\vev}[1]{\langle #1 \rangle}

\newcommand{\vphi}{\varphi}

\newcommand{\cH}{{\cal H}}
\newcommand{\cL}{{\cal L}}

\newcommand{\cW}{{\cal W}}

\newcommand{\cS}{{\cal S}}

\newcommand{\bbR}{\mathbb{R}}
\newcommand{\bbZ}{\mathbb{Z}}
\newcommand{\bb}[1]{\mathbb{#1}}

\newcommand{\shl}[1]{{#1}_{\hat{l}}\,}
\newcommand{\lh}[1]{{#1}^{(\hat{l})}}
\input epsf
\begin{document}  

\begin{center}{\Large\bf Wavelet field decomposition and UV `opaqueness' }\\[2cm] 
{E. T. Tomboulis\footnote{\sf e-mail: tomboulis@physics.ucla.edu}
}\\
{\em Mani L. Bhaumik Institute for Theoretical Physics\\
Department of Physics and Astronomy, UCLA, Los Angeles, 
CA 90095-1547} 
\end{center}
\vspace{1cm}

\begin{center}{\Large\bf Abstract}\end{center}  
A large body of work over several decades indicates that, in the presence of gravitational interactions, there is  loss of localization resolution within a fundamental ( $\sim$ Planck) length scale $\ell$. We develop a general formalism based on wavelet decomposition of fields that takes this UV `opaqueness' into account in a natural and mathematically well-defined manner. This is done by requiring fields in a local Lagrangian to be expandable in only the scaling parts of a (complete or, in a more general version, partial) wavelet Multi-Resolution Analysis. This delocalizes the interactions, now mediated through the opaque regions, inside which they are rapidly decaying. The opaque regions themselves are capable of discrete excitations of $\sim 1/\ell$ spacing.   
The resulting effective Feynman rules, which  give UV regulated and (perturbatively) unitary physical amplitudes, resemble those of string field theory.

\vfill
\pagebreak

\section{Introduction  \label{Intro}} 
\setcounter{equation}{0}
\setcounter{Roman}{0}

There is a large body of well-known arguments pointing to  
a fundamental limitation in the localization resolution any conceivable experiment can achieve. At an elementary level, this first arises when taking onto account the fact that, in any localization measurement, 
both the probe and the probed particle carry energy and hence gravitate. This effect, normally utterly  negligible for the derivation of the standard uncertainty relations by the familiar physical considerations (`Heisenberg microscope'), becomes dominant when it comes to localization within scales of the order of Planck length. Even for  optimal probes (photons),  
avoidance of formation of a horizon sets a fundamental limit in resolution of the order of some fundamental scale (Planck length).  Arguments of this type go, remarkably, back  to the 1930s. In more recent decades, this lack of resolution, or `UV opaqueness', has been discussed from several field- and string-theoretic points of view and explored in calculations of scattering at trans-Planckian energies. For a review, including some of the early interesting history, and extensive references see \cite{H}.  

Such a limitation implies that all interaction vertices must acquire some nonlocality characterized by this fundamental scale.  
Nonlocal vertices occur in a variety of field theory models and in string field theory. 
Such nonlocal field theories, both in their own right or as models of string field theory Feynman rules, have recently been studied, in particular, with respect to unitarity, analyticity and causality properties of their amplitudes \cite{CT1}, \cite{PS}, \cite{T1}.  In these theories interactions within a fundamental scale become smeared and `soft' (rapidly decaying along the Euclidean directions) with resulting loss of resolution.

This paper presents  a framework within field theory for incorporating loss of resolution inside some scale $\ell$ in a natural manner. This framework is based on wavelet decomposition of fields. 
The theory of wavelets may be viewed in the wider context of the development, in recent decades, of the subject of `atomic decompositions', the decomposition of function spaces in `atoms' or `molecules', cf., e.g.,  \cite{M}.  Wavelets, more specifically,  
implement a Multi-Resolution Analysis (MRA) of function spaces \cite{D}, \cite{M}.\footnote{The literature on wavelets is by now vast. \cite{D}, however, remains the classic presentation of the mathematical theory of wavelets. Another authoritative text, at a somewhat higher level, is \cite{M}.} 
This consists of decomposing a function space, say $L^2$, in subspaces, each of which refers to a certain resolution range (in time, space, spacetime, or any other `length' variables characteristic to the system at hand). 
The precise definition of MRA and the basic properties of the resulting wavelet expansions are given in the Appendix. Briefly, a wavelet expansion of a field $\phi$ in $\bb{R}^d$ is of the form 
\beq  
\phi(x) = \sum_n \phi_n \sigma_n(x) + \sum_A \phi_A \upsilon_A(x)    \; . \label{0w-fexp}  
\eeq
Here $\sigma_n(x$) 
are the set of scaling fields obtained by dilation to the scale $\ell$ and discrete translations $n\in \bb{Z}^d$ of a mother scaling function $\sigma(x)$ (cf. (\ref{bas1}) below). The wavelets $\upsilon_A$, enumerated by the multi-index $A=(m,n,q)$, are obtained by dilation to length scales $\ell^m$, $m \in \bb{Z}^+$,  and discrete translations $n \in \bb{Z}^d$ of the mother wavelet $\upsilon(x)$ associated with $\sigma(x)$; there must be $2^d-1$ such wavelets for each $m,n$ enumerated by $q$ (cf. (\ref{bas2}) - (\ref{bas3}) below).  
The scaling part represents resolution at all length scales from infinity to $\ell$;  whereas, the wavelet parts provide successively finer resolution from $\ell$ down to arbitrarily small scales. It may be noted that the scaling part itself may be expanded in wavelets pertaining to longer length scales up to some  $\ell^\prime > \ell$ (including $\ell^\prime \to \infty$, cf. Appendix); but, except for some special purposes, (\ref{0w-fexp}) is the standard expansion for some $\ell$ appropriate to the system under consideration.  

The set $\{\sigma_n, \upsilon_A\}$ is a complete, orthonormal basis for the function space, which is decomposed in a direct sum of the subspaces spanned by the corresponding elements of this set (cf. Appendix). This implies that working with  particular subsets of components $\phi_n, \phi_A$ can be done by using suitable projection operators acting on the full field $\phi$. This is very convenient for setting up a formalism suitable 
for weak coupling perturbation theory as will be seen in the following sections. 

The fact that such separation of length scales can be achieved in an {\it orthogonal} decomposition is astonishing and the existence of such orthonormal wavelet bases mathematically highly non-trivial. It was the explicit construction of such sets in the 1980s that led to the subsequent explosive development and application of wavelets.\footnote{Orthogonal wavelet decompositions are a  very non-trivial special case of the more general mathematical theory of {\it frames} \cite{D}. Frames, roughly speaking, are redundant (over-complete) sets of linearly dependent vectors used to expand vectors in (the concept is more subtle for infinite dimensional spaces). 
The extreme case of a frame is the continuous wavelet transform, which replaces the discrete $n$, $m$ in (\ref{0w-fexp}) by continuous variables resulting in an infinite-degenerate representation of functions: every component is linearly dependent on all the others. Such redundant descriptions can be very useful in engineering in assuring robustness of, say, reconstruction of a signal; but they appear very ill-suited for unambiguously re-expressing the path integral in field theory.} 

Though by now wavelets are ubiquitous in all fields of engineering and many physical and medical science areas and beyond, their use in field theory has been quite limited. This is somewhat surprising since they would appear to be tailor-made for real space renormalization group application. Wilsonian RG was, in fact, one of several  precursor body of ideas that led to wavelets.\footnote{The `Wilson basis' is  frequently referred to in the mathematical literature leading to wavelets.}  This limited use has indeed been mostly in RG and  lattice gauge theory within a  constructive approach \cite{F}, or a more general lattice field framework \cite{N1}, \cite{N2},  or numerical and MC implementation \cite{Hl}, \cite{BS}, \cite{DMcN}. More recently, the use of wavelets in field theory has been advocated in \cite{BP}, \cite{P}.

In this paper, the wavelet  framework of separating length scales is used to implement sequestering of scales smaller than a length $\ell$. The basic, simple idea is that, given a field theory Lagrangian $\cL(\vphi)$ with local interactions, the fact that the fields $\vphi$, or any observable constructed out of them, cannot actually be resolved beyond a fundamental scale $\ell$, implies that their wavelet decomposition cannot contain wavelet components probing lengths shorter than $\ell$. Since we can always adjust the decomposition to have $\ell$ as the characteristic length of the scaling space, this means that $\vphi$ is expandable in only scaling components. 
This has a number of implications whose derivation follows in a  straightforward and mathematically well-defined manner. Regions of length order $\ell$ become UV opaque, i.e.,  suffer loss of resolution, implying a resulting delocalization of interactions. The opaque regions appear as Euclidean spacetime `atoms' mediating this delocalization. They can emit and absorb momenta in discrete units of $1/\ell$ as a consequence of the translation invariance of the theory. It must be stressed here that no explicit cutoffs of any type are employed. Fields are defined for all $x\in \bb{R}^d$; and, correspondingly, momenta vary unrestricted up to infinity.  
The formulation is in Euclidean field theory within an overall $S$-matrix approach.  Minkowski amplitudes must be defined by analytic  continuation of external momenta of the Euclidean amplitudes. 

Only global symmetries are considered in this paper. 
The extension of the formalism required to accommodate local symmetries will be considered in a separate treatment.

An outline of the paper is as follows. 
The wavelet expansion of fields in Euclidean space $\bb{R}^d$ is detailed in section \ref{WD}. Additional mathematical background related to such expansions is included in the Appendix, where the general framework of MRA is presented.

In section \ref{UVs} the basic formalism of field theory models with fields expandable in only scaling components is developed. The crucial requirements here is that scale and wavelet mother functions possess Fourier transforms of sufficiently rapid decay along the Euclidean axis and be entire functions on $\bb{C}^d$. 
The resulting delocalized vertices are then given by entire functions in momentum space.  
This ensures perturbative unitarity and appropriate singularity structure as discussed in detail in \cite{PS}, \cite{CT1}. Furthermore, these delocalized vertices imply UV finiteness for a wide variety of interactions.

In section \ref{GF}, abstracting from the construction in the preceding sections, another formulation is developed, which, though wavelet-inspired, no longer relies on having a complete MRA and, therefore, on any specific wavelet realization. The only elements retained is a projection operator onto a subspace of the `coarse' part of fields, 
of limited resolution within a scale $\ell$, and the orthogonal complement onto the rest of the field function space. 
This provides a wider and more flexible framework, which evades certain technical issues with higher dimensional wavelets as described in the text, and allows implementation in essentially any physically sensible field theory model. The resulting Feynman rules are of the type encountered in string field theory.

Section \ref{DC} contains some concluding further remarks and discussion of open issues.

We use standard mathematical notations. In particular, $\widehat{f}$ stands for the Fourier transform of $f$;  
and $\vev{f,g} = \int d^dx \; \bar{f}(x) g(x)$ for scalar products with the bar denoting complex conjugation.

\section{Wavelet decomposition of fields   \label{WD}} 
\setcounter{equation}{0}
\setcounter{Roman}{0}

\subsection{The general wavelet expansion} 
We fix some UV length scale $\ell$, which here may be naturally taken to be of the order of Planck length or some unification scale.  
With $\hat{\ell}$ denoting this scale in dimensionless units,\footnote{We could, of course, by choice of units, set $\ell=1$, and $\hat{l}=0$, thus simplifying notations, but it is preferable to keep reference to this UV scale 
explicit. Expressing $\ell$ in powers of $2$ is the standard practice in wavelet theory.} we set 
\beq 
\hat{\ell} = 2^{-\hat{l}}  \;  
\label{lscale1}
\eeq
with integer $\hat{l}$. A Multi-Resolution Analysis consists of a sequence of orthogonal spaces of increasingly finer resolution starting with the scaling space referring to a given scale (see Appendix). So here we take the scaling space,  denoted $V_{\hat{l}}$, to refer to the scale $\ell$. 
This means that, given a mother scaling function $\sigma(x)$  and corresponding $2^d-1$ mother wavelet functions $\upsilon^q(x)$ on $\bbR^d$, the basis set is given by 
\bal
\sigma_{\hat{l}n}(x) & = 2^{d\hat{l}/2} \sigma(2^{\hat{l}} x - bn)  
  \label{bas1}\\
\upsilon^q_{m n}(x) & = 2^{d m/2} \upsilon^q(2^m x - bn)    \;  \label{bas2}\\
 \mbox{with}\qquad  \qquad  x\in \bbR^d \; , \qquad   n \in \mathbb{Z}^d \;,  &   \qquad  \hl  \leq m \in \bbZ  \; , \qquad  1\leq q \leq 2^d-1    \; \qquad 0< b \, \in  \bbR^+ \; . \label{bas3}       
\end{align} 
$b$ is the translation parameter.  
The scaling set $\sigma_{\hl n}(x)$ constitute a basis in $V_{\hl}$, 
whereas, for each $m$, the set $\upsilon^q_{mn}(x)$ constitute a basis for the $m$-th resolution space $W_m$. The spaces are mutually orthogonal and one has the orthogonality relations:  
\bal
\int d^dx\;  \overline{\sigma}_{\hl n}(x) \sigma_{\hl k}(x)  & = \delta_{nk}   \label{orel1}     \\
\int d^dx  \;  \overline{\sigma}_{\hl n}(x) \upsilon^q_{m k}(x) & =0 \; , \qquad \qquad  m\geq \hl  \label{orel2}   \\ 
\int \overline{\upsilon}^q_{mn}(x) \upsilon^{q^\prime}_{m^\prime k}(x)  & =\delta_{m m^\prime}  \delta_{nk} \delta_{q q^\prime}   \; . \label{orel3}  
\end{align} 
The mother functions $\sigma(x)$ and $\upsilon(x)$ are generally well-localized around the origin within a length of order $\ell$; they may, in particular, be of compact support. 
Physical requirements dictate that the FT  $\widehat{\sigma}(k)$ be of sufficiently rapid decay along the Euclidean momentum axis and an entire function on $\bb{C}^d$. 
 .

A field configuration $\phi(x)$ on $\bbR^d$ has the wavelet expansion  
\beq 
\phi(x) =   \sum_n \phi_n \sigma_{\hl n}(x) + \sum_{q, m,n} \phi_{mn}^q  \upsilon^q_{mn}(x)    \; ,  \label{fexp1}
\eeq
with 
\beq 
\phi_n = \vev{\sigma_{\hl n}, \phi} \;, \quad \phi^q_{m n} = \vev{\upsilon^q_{mn}, \phi}  
\label{fexp2} 
\eeq
and  summations over $n, m, q$ as specified in (\ref{bas3}).

As explained in the Appendix, (\ref{fexp1}) is a decomposition in successively finer resolutions. 
The scaling field part $\sum_n \phi_n \sigma_{\hl n}(x)$ may be viewed as the `coarse' part of the field, which is insensitive to (smeared over)  regions of scale $\ell$. It thus represents all features of $\phi$ down to scale $\ell$.  The wavelet parts  probe inside such regions down to arbitrarily small scales with successively finer resolution: the $\upsilon^q_{mn}$ wavelets terms probe scales of order $\ell^m$, $m\geq 1$.  
The remarkable, and very nontrivial, fact here is that this separation of scales is accomplished in an orthogonal exact decomposition.

The basis (\ref{bas1}) - (\ref{bas2}),  
being orthonormal and complete, provides a complete resolution of the identity. The set of coefficients $\{ \phi_n, \phi^q_{mn} \} $ gives then a discrete representation  of $\phi(x)$ with the coefficients as the dynamical degrees of freedom. This representation, with integration over $\phi$ replaced by integration over the infinite set $\{ \phi_n, \phi^q_{mn}\}$ in the functional integral, is the natural one in applications such as real space renormalization group \cite{F},  \cite{BP}. For our purposes here, however, it will be often more convenient to revert to use of $\phi(x)$ as dynamical variable.  The connection is simply provided by the projection operators onto the subspaces $V_{\hat{l}}$ and $W_{m}$ of the decomposition 
(\ref{fexp1}) given by the $d$-dimensional version of (\ref{Pproj}) and (\ref{Qproj}), i.e., 
\bal 
P_{\hl }(x,y) &  =  \sum_n \sigma_{\hl n}(x) \overline{\sigma}_{\hl n}(y)  \label{dPproj} \\
Q_m(x,y)& =\sum_{q, n} \upsilon^q_{mn}(x) \overline{\upsilon}^q_{mn}(y)    \label{dQproj}       \; .          
\end{align} 

In terms of these (\ref{fexp1}) assumes the form 
\bal
\phi(x) &=\int d^dy \, P_{\hl}\, (x,y) \phi(y)  +\sum_{m}   \int d^dy \, Q_m(x,y) \phi(y)  \\
& = (P_{\hl}\, \phi)(x)  + \sum_m (Q_m \phi)(x)     
\; .   \label{fexp3}
\end{align}
Introducing the projection to the direct sum of all $W_m$ subspaces 
\beq 
\shl{Q}(x,y) = \sum_{m\geq {\hl}} Q_m(x,y)     \label{dQprojtotal} 
\eeq
one has 
\beq
\shl{P} + \shl{Q} = 1   \;   \label{P+Qproj} 
\eeq
and, by the orthogonality relations (\ref{orel1}) -(\ref{orel3}), one indeed has $(P_{\hl}^2)(x,y) = P_{\hl}(x,y)$, 
$(Q_{\hl}^2)(x,y) = Q_{\hl}(x,y)$, and $(P_{\hl}Q_{\hl})(x,y) = 0$. 
The projections $\shl{P}$ and $\shl{Q}=1-\shl{P}$ then decompose the space of field configurations $\cS$ into two orthogonal subspaces: the subspace $\shl{V}$ of fields representing features down to scale $\ell$; and the subspace $\shl{\cW}=\bigoplus_{m\geq \hl} W_m$ of fields that can represent features from $\ell$ down to arbitrarily small length scales.  Thus, from (\ref{fexp3}): 
\beq
\phi(x) = (\shl{P} \phi)(x) +( \shl{Q}\phi)(x)    \; . \label{P+Qexp}
\eeq

In (\ref{fexp1}) $\phi$ is a scalar field and the wavelet basis functions $\sigma(x)$, $\upsilon(x)$ are assigned scalar transformation properties under $SO(d)$ transformation. For tensor fields $\phi^\alpha$, where $\alpha$ is a generic tensorial index, the tensorial properties are carried by the expansion coefficients. Thus, under an $SO(d)$ rotation $\Lambda$ one has 
\bal
\phi^{\prime \alpha}(x) & = D(\Lambda)^\alpha_\beta \, \phi^\beta (\Lambda^{-1} x)  \nonumber \\
& =   \sum_n D(\Lambda)^\alpha_\beta\, \phi^\beta _n \sigma_{\hl n}(\Lambda^{-1}x) + \sum_{q, m,n} D(\Lambda)^\alpha_\beta\, \phi_{mn}^{q\beta}  \upsilon^q_{mn}(\Lambda^{-1}x)    \; ,  \label{tfexp1}
\end{align} 
where $D(\Lambda)$ denotes the transformation matrix in the appropriate tensor representation.

\subsection{Basis functions} 

To implement this general framework in actual computations we need 
an explicit realization of the basis functions $\sigma_n(x)$ and $\upsilon^q_{mn}(x)$.  
In practice, the only widely available way to do this in $d>1$ is to form the $d$-dimensional basis as the  tensor product of $1$-dimensional bases (see remarks in Appendix). 

Denoting the scaling and wavelet functions in $d=1$ by $\rs(x)$ and $\rw(x)$, the $1$-dimensional basis is given by the $d=1$ version of (\ref{bas1}) - (\ref{bas3}) (Cf. Appendix). 
If we now take our orthonormal Euclidean $d$-dimensional basis   
(\ref{bas1}) - (\ref{bas3}) to be a tensor product basis formed out of 1-dimensional bases, it can be conveniently expressed as follows. 
Let $E$ denote the set of $2^d-1$ sequences $\{q_1, q_2, \ldots, q_d\}$ of $0$s and $1$s excluding the sequence $\{0, 0,\ldots,0\}$ of only zeroes. Let 
\beq 
\rw^{(q_\mu)}(x) \equiv \rw(x) \qquad   \mbox{if} \quad q_\mu=1 \; ;  \qquad 
\rw^{(q_\mu)}(x) \equiv\rs(x) \qquad   \mbox{if} \quad q_\mu=0   \;.   \label{Esetdef}
\eeq
Then the tensor product realization of the $d$-dimensional basis (\ref{bas1}) - (\ref{bas3}) is: 
\bal 
\sigma_{ \hl n}(x) & = \prod_{\mu=1}^d \rs_{\hat{l} n_\mu}(x_\mu) = 2^{d\,{\hl}/2} \prod_{\mu=1}^d \rs(2^{\hl} x_\mu - bn_\mu) \; ,   \qquad \quad    n_\mu \in \bbZ \label{tens1} \\
\upsilon^q_{mn}(x)  & =  2^{d m/2} \prod_{\mu=1}^d \rw^{(q_\mu)}(2^m x_\mu - bn_\mu)   \; ,   \qquad m\geq \hat{l} \; ,  \quad m, n_\mu \in \bbZ\;,  \quad q\in E \; .  \label{tens2}       
\end{align}
Note that the tensor product realization makes it obvious why we need $2^d-1$ different wavelets for a given scale $m$ in $d$-dimensions. But this holds in general, i.e., when basis (\ref{bas1}) - (\ref{bas3}) is not necessarily a tensor product basis.\footnote{For the existence proof of the general basis (\ref{bas1}) - (\ref{bas3}) see \cite{M}, Ch. 3.}

Scaling and wavelet functions are generically constructed to be well localized, though not necessarily of compact support. 
To be definite, we may employ `minimal' Daubechies wavelets (sometimes denoted by ${}_{\ssc N}\rs(x)$, ${}_{\ssc N}\rw(x)$), which are of compact support \cite{D}. 
They are characterized by a positive integer $N$.\footnote{There is no explicit analytic formula for Daubechies wavelets; their existence and construction to arbitrary accuracy is obtained through the `cascade algorithm'. $N$ refers to the fact that the mother scaling field $\rs(x)$ has $2N$ nonvanishing filter coefficients, cf. \cite{D}.} 
$N=1$ gives the Haar wavelet. Both $\rs(x)$ and the mother wavelet $\rw(x)$ have compact support on the interval  $[0, (2N-1)]$, hence support width $(2N-1)$. Hence $s_{\hat{l} n}(x)$ has support on $x \in [2^{-\hat{l}} bn, 2^{-\hat{l}}(bn + 2N-1)]$; and 
$w_{mn}(x)$ has support on $x \in [2^{-m}bn, 2^{-m}(bn + 2N-1)]$, i.e., support width $2^{-m}(2N-1)$ with $m\geq \hat{l}$, thus providing successively finer resolution.

With $\rs(x)$ and $\rw(x)$ being functions of compact support, their Fourier transforms $\hat{\rs}(\xi)$ and $\hat{\rw}(\xi)$ are $C^\infty$ functions, and, in fact, entire functions for $\xi \in \mathbb{C}$.  This is a fundamental requirement for application to quantum field theory. 
For real (Euclidean) $\xi$, $\hat{\rs}(\xi)$ has the asymptotic decay (\cite{D}, Chapter 7): 
\beq
|\,\hat{\rs}(\xi) | \leq C (1+ |\xi|)^{- 1 - (N-1)r}     \; , \label{DFsfdecay}
\eeq
where $r= (1 - \frac{\ln 3}{2\ln 2}) \simeq 0.21$.

From (\ref{tens1}) - (\ref{tens2}) we see that the $d$-dimensional $\sigma_{\hl n}(x)$ has support on $x \in \prod_\mu[2^{-\hat{l}} b n_\mu, 2^{-\hat{l}}(b n_\mu + 2N-1)]$, i.e., within a $d$-dim box of side length $(2N-1)2^{-\hl}$; whereas the wavelets 
$\upsilon^q_{mn}(x)$ have support  
in $d$-dim boxes with some sides of length $(2N-1)2^{-\hl}$ and some of length  $2^{-m}(2N-1)$, with $m \geq \hat{l}$. 
When $q_\mu=1$ for all $\mu$ all sides have the latter length.   
The part of the field expanded in the scaling fields $\sigma_{\hl n}$ represents all features of $\phi$ down to scales $(2N-1) 2^{-\hat{l}}$. Features down to scale 
$(2N-1) 2^{-(\hl + 1)}$ that cannot be represented on scale $(2N-1) 2^{-\hl}$ are represented by the wavelets $\upsilon^q_{mn}$ terms with $m= \hl$; features at scale $(2N-1) 2^{-(\hl + 2)}$
that cannot be represented on scale $(2N-1) 2^{-(\hl+1)}$ are represented by the wavelets $\upsilon^q_{mn}$ terms with $m= \hl+ 1$, and so on. 
In the following, as it is customary, we set the translation parameter $b$ equal to one.

The use of a tensor product basis formed from 1-dimensional basis functions, however, presents a technical problem for applications in field theory as envisioned here. Such a basis selects a particular frame so that manifest $SO(d)$ invariance is lost. One would like to have radial MRA basis functions, i.e., start from radial scale and wavelets mother functions $\sigma(x) = \sigma(|x|)$, $\upsilon(x)= \upsilon(|x|)$, where $|x|$ denotes the usual Euclidean norm in $\bb{R}^d$, so that basis functions are manifestly rotationally invariant. Radial MRA wavelets have been constructed for $d=3$ \cite{RR}, \cite{CP}, though such construction are rather involved. There are apparently no such constructions available for $d\geq 4$ though. One may use general radial basis theory to construct some wavelets systems \cite{B}, so this remains an open issue.  
The field theory models in the next section are constructed for any general $d$-dimensional basis (\ref{bas1})-(\ref{bas3}); but explicit use of a specific MRA basis faces this technical issue.    
The generalized formulation in section 4, on the other hand, by going outside a complete MRA, 
allows use of radial functions from the outset.

\section{Field theory model with limited UV resolution   \label{UVs}} 
\setcounter{equation}{0}
\setcounter{Roman}{0}

As alluded to in the introduction the basic idea is very simple.  
Wavelet decomposition like (\ref{fexp1}) allows one to accommodate different spatial resolution levels.

Consider the simplest case of real scalar fields. 
Let 
\beq
\vphi(x) = \sum_n \vphi_n \sigma_{\hl n}(x)  \; .   \label{sf1}
\eeq
and  
\beq
\chi(x) = \sum_{q,m,n} \chi^q_{mn} \upsilon^q_{mn}(x)       \; . \label{wf1}
\eeq
The sets $\{\vphi_n\}, \{\chi^q_{mn}\}$ are the dynamical degrees of freedom. 
$\vphi \,\epsilon \, \shl{V} $ is insensitive to 
(compact) regions of scale $\ell$, whereas $\chi(x)\epsilon \shl{\cW}$ is built of wavelets probing all scales shorter than $\ell$.  
We then define the model given by the Euclidean action 
\beq
S =  \int d^d x\; \Big( \frac{1}{2} \vphi  K \vphi   
+  \cL_I(\vphi)  \Big)   \;  ,   \label{S1}
\eeq
where $\cL_I$ is some local interaction Lagrangian and 
\beq
K(\partial) = \big( - \Delta + m^2\big)     \;    \label{kin1}
\eeq
with $\Delta = \delta^{\mu\nu} \partial_\mu\partial_\nu$.  This implements the idea that regions of length scale $\ell$  become `opaque' and cannot be probed by interactions so as to achieve finer resolution. Here these regions are the support of the scaling fields $\sigma_{\hl n}$, assumed to be well-localized  around $2^{-\hl} n$.

Quantization is performed via the Euclidean path integral: 
\beq
Z[J] = \int [D\vphi] \;  \exp{\D\Big\{- \int \big( \frac{1}{2} \vphi K \vphi  +  \cL_I(\vphi) + J\vphi   \big) \Big\}} 
 \; .  \label{PI1} 
\eeq
It is important to note that working with $\vphi$ and  $\chi$, as, e.g., in the purely $\vphi$-dependent action (\ref{S1}), 
means that $\vphi_n$ and $\chi^q_{mn}$ are the independent degrees of freedom. 
Integration over $\vphi$ and $\chi$ is defined with measures $[D\vphi]$ and  $[D\chi]$, where 
\bal
[D\vphi] &\equiv \prod_{n} d\vphi_n   \label{meas1}\\
[D\chi] & \equiv \prod_{q,m,n} d\chi^q_{mn}   \;  , \label{meas2}
\end{align}
and  the action expressed in terms of $\{\vphi_n, \chi^q_{mn}\}$; e.g., for the $\vphi$ kinetic term in (\ref{S1}): 
\beq
 \frac{1}{2} \vphi K \vphi  = \frac{1}{2}\sum_{n,n^\prime} \vphi_n K_{nn} \vphi_{n^\prime}  \;, \qquad 
K_{n n^\prime}= \vev{\sigma_{\hl n}, K \sigma_{\hl n^\prime} }  \; ,  \label{S1a}
\eeq
and so on for the other terms in the action.

\begin{figure}[ht]
\begin{center}
\includegraphics[width=15cm]{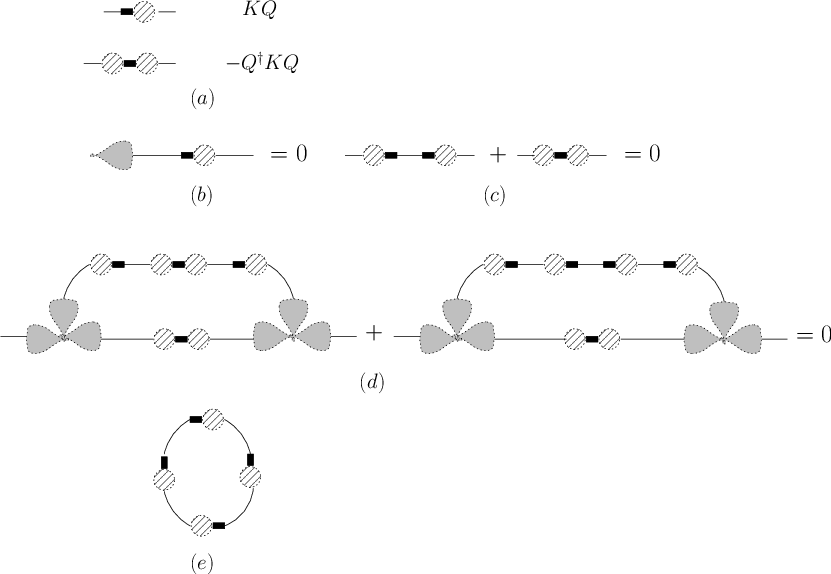}
\end{center}
\caption{Figure 1\ (a) The two $2$-point vertices in (\ref{S3}); (b) - (c) the two basic cancelation mechanisms; (d) an example of cancelation via (c) in a generic loop graph; (e) residual vacuum graphs - see text. \label{wfdF1}}
\end{figure}

 In general, as already noted, working with the wavelet decomposition coefficients $\{\vphi_n, \chi^q_{mn}\}$,   
 though natural in the context of Wilsonian RG (block spinning) and similar nonperturbative constructions, cf. \cite{BP}, \cite{F}, is not well-suited for perturbation considerations.  
To be able to conveniently work with (\ref{S1})-(\ref{PI1}) within weak coupling perturbation theory we first include in (\ref{PI1}) also integration over the fields $\chi\in \mathcal{W}_{\hl}$; the fields $\chi$ do not appear in the action and are thus decoupled.
Now, $\vphi$ and $\chi$ are independent, in fact orthogonal, fields. We may add them to form a field $\phi$ containing components at all length scales (c.f. (\ref{fexp1})):
\beq 
\phi = \vphi + \chi \; , \qquad P_{\hl}\,\phi = \vphi \;, \quad Q_{\hl}\, \phi  = \chi    \label{fexp4} 
\eeq
with components 
\beq 
\phi_n = \vev{\sigma_{\hl n}, \phi}=\vphi_n  \;, \quad \phi^q_{m,n} = \vev{\upsilon^q_{mn}, \phi}  =\chi^q_{mn} \;.
\label{fexp5} 
\eeq
Thus the $[D\vphi]$ measure in (\ref{PI1}) is extended to: 
\beq 
[D\vphi] [D\chi] = [D\phi]  =\prod_x \, d\phi(x)   \; ,  \label{meas3}
\eeq 
i.e., the usual formal functional measure over $\phi$. In terms of $\phi$ 
the action (\ref{S1}) assumes the form: 
\beq
\int d^d x\; \Big( \frac{1}{2}\phi P^\dagger_{\hl}K P_{\hl}\, \phi  
+  \cL_I(\shl{P}\phi)  \Big)    \; . \label{S2}
\eeq
It should be noted that the action, in the form (\ref{S1}), is local in terms of the coarse field $\vphi$, but, in the form (\ref{S2}), is nonlocal in terms of $\phi$.\footnote{Though an elementary point, it is still perhaps worth remarking  that (\ref{S2}) is certainly {\it not} a field redefinition of the usual local action  
\[  \int d^d x\; \Big( \frac{1}{2} \phi K  \phi  
+  \cL_I(\phi)  \Big)   \; . \]
Indeed, $\phi \to P\phi=\vphi$  would not be a legitimate change of variables in the functional measure since ${\rm Det} P=0$ as $P$, being  a projection, has  nontrivial null space. The point is, of course, obvious in terms of the orthogonal compliments $\vphi$ and $\chi$.}  
We next rewrite   
the quadratic kinetic term  by inserting  
\beq
\vphi = \shl{P} \phi= \phi - \chi  = \phi- \shl{Q} \phi\; .  \label{S3-fields} 
\eeq
One obtains  
\beq
S= \int d^d x\; \Big( \frac{1}{2} \phi K \phi   -  \phi K \shl{Q} \phi 
+ \frac{1}{2} \phi\, Q^\dagger_{\hl} K \shl{Q}\phi +  \cL_I(\shl{P} \phi) + J\shl{P}\phi  \Big)   \; . \label{S3}
\eeq 
We now treat the first term in (\ref{S3}) as defining the propagator and all others as interactions. Thus, in addition to the original 
interaction vertices from $\cL_I(\shl{P}\phi)$ and external source vertices $J\shl{P}\phi$, we have  the 2-point interaction vertices $K\shl{Q}$ and $Q^\dagger_{\hl} K\shl{Q}$ as shown in Fig.\ref{wfdF1}(a). All contributions from these additional 2-point vertices, however, cancel by the two simple mechanisms depicted in Fig.\ref{wfdF1}(b) - (c): 
(i) The propagator $K^{-1}$ connected to an insertion of a 2-point vertex $ K\shl{Q}$ in a line in a graph ends at a vertex from $\cL_I$ or an external source (Fig.\ref{wfdF1}(b)). This generates the structure 
\[  \shl{P}K^{-1} K\shl{Q}K^{-1} =\shl{P}\shl{Q}K^{-1}  = 0    \; , \] 
 since $\shl{P}$ and $\shl{Q}$ are orthogonal projections. (ii) Insertion of two $KQ$ vertices connected as shown in Fig. \ref{wfdF1}(c) in a line in a graph is cancelled by the insertion of one  $Q^\dagger K Q$ vertex in the same line in an otherwise identical graph, since 
 \[ K^{-1} \big[Q^\dagger K K^{-1} KQ  + (- Q^\dagger K Q)\big] K^{-1}  = 0 \; . \]
It is easily seen that (i) and (ii) are present  no matter how many insertions, and in what order, of 
the 2-point vertices of Fig. \ref{wfdF1}(a) occur in any line of a (tree or loop)  graph: either orthogonality with a $\cL_I$ vertex, or cancellation via Fig. \ref{wfdF1}(c) in pairs of otherwise identical graphs, ensure cancelation of all contributions from 2-point vertices in any amplitude or expectation. An example is shown in  Fig. \ref{wfdF1}(d).  
Among vacuum graphs, however, there is one exception. There is no cancelation for vacuum graphs of the type shown in Fig. \ref{wfdF1}(e). These, however, being disconnected vacuum graphs, cancel in any expectation; or, equivalently, they may be cancelled by a redefinition of the functional measure by a constant ( field independent) factor.\footnote{Similar cancellations of the type in  Fig. \ref{wfdF1}(b)-(c) are exhibited for general (nonlocal) changes of variables in \cite{tHV1}. The additional element used here is the orthogonality projections in the interaction vertices, which is special to the theories considered here.} 
There is, of course, no surprise here since these cancellations simply reflect the fact that the $\chi$ fields are decoupled in (\ref{S1}). 
In the computation of any amplitude then one may simply drop the 2-point vertices in the action  (\ref{S3}).  
The resulting effective rules for the action (\ref{S3}), now with measure (\ref{meas3})  in (\ref{PI1}), are thus the same as the Feynman rules as for the action:
\beq
S =  \int d^d x\; \Big( \frac{1}{2} \phi K \phi   
+  \cL_I(P_{\hl}\,\phi)  \Big)   + J\shl{P}\phi  \;  .     \label{S4}
\eeq
The form of the action (\ref{S4}) is well-suited for weak coupling perturbation theory.    
In (\ref{S4}) one has the normal free kinetic part but the vertices involve the projection kernels in 
$(P\phi)(x)$ and, thus, are nonlocal. (\ref{S4}) is, in fact, precisely of the general `canonical' form
that has recently been studied in models of nonlocal field  and string field  theory Feynman rules. That is, one has the usual (Euclidean) propagator 
\beq
\Delta(k)  = K^{-1}(k) = \frac{1}{k^2 + m^2}          \label{prop1}
\eeq
and nonlocal vertices. 
Here, however, the structure of the nonlocal vertices arises in a definite way as a result of the projection onto the scaling part of the wavelet decomposition.  

Consider an  interaction term in (\ref{S1}), respectively in (\ref{S2}), (\ref{S4})  of $N$ fields:
\beq
S_N 
= \frac{1}{N!} \int d^d x \;  \vphi(x)^N  = \frac{1}{N!} \int d^d x \;  (P_{\hl}\, \phi)^N(x)      \; .      \label{V1}
\eeq
In terms of $\phi$, this represents the local interaction, at the point $x$, of $N$ delocalized fields $(P\phi)(x)$. 
(\ref{V1}), written in momentum space, is seen after a short computation using (\ref{dPproj}) to be given by: 
\bal 
S_N & = \int \prod_{i=1}^N \frac{d^d k_i}{(2\pi)^d} \sum_{l_1, \ldots,l_N} (2\pi) \delta^{(d)}\Big( \sum_{j=1}^N \big(k_j + 
2^{\hat{l}} 2\pi l_j\big)\Big)      \nonumber    \\
& \qquad \qquad \cdot  \frac{1}{N!} \,  \lh{V}_N(k_1, \ldots, k_N, l_1, \ldots, l_N) \; \phi(k_1) \cdots \phi(k_N)   \; ,    \label{V2}
\end{align}  
with vertex factor  
\beq
 \lh{V}_N(k_1, \ldots, k_N, l_1, \ldots, l_N) = \prod_{i=1}^N \widehat{\sigma} \big(2^{-\hl} k_i +  2\pi l_i\big) \, \overline{\widehat{\sigma}}(2^{-\hl}k_i)  \; .   \label{V3}
\eeq
Here $\widehat{\sigma}(q)$ denotes the FT of the mother scaling function $\sigma(x)$. 
The summation in (\ref{V2}) is over integer component vectors $l_i = (l_{i1}, \ldots , l_{i d}) \in \bbZ^d$.  

\begin{figure}[ht]
\begin{center}
\includegraphics[width=5cm]{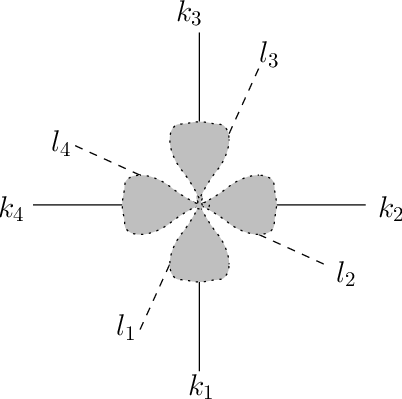}
\end{center}
\caption{ The vertex given by eq. (\ref{V2})-(\ref{V3}) depicted for $N=4$. The shaded blobs represent the opaque regions.     \label{wfdF2}}
\end{figure}

The vertex (\ref{V2}) - (\ref{V3}) (Fig. \ref{wfdF2}) has an interesting structure. It represents the interaction of $N$ `boxes' (regions of size of the support of $\sigma_{\hl n}(x)$, i.e., of linear size of order $\ell$, as described in section \ref{WD}) interacting at the spacetime point $x$ by (\ref{V1}). They may be viewed as Euclidean spacetime `atoms'.  
There are $N$ incoming external particle momenta $\{k_i\}$. The boxes (atoms) can emit or absorb momentum in discrete units of $2^{\hat{l}} 2\pi $.  
This is a consequence of the MRA discretization of the translation operation on the scaling function (cf. (\ref{bas1})). This translation fixes the location of the box and, upon quantization, acts as a collective coordinate with conjugate momentum. This restores translation invariance, hence overall  momentum conservation. This conservation must, of course, be there since the theory, in particular, the interaction (\ref{V1}), is translation invariant.  The vertex (\ref{V3}) is depicted in Fig 1. 

The gap to the first ($l_\mu=1$) box excitation is $2^{\hat{l}} 2\pi$, so it becomes relevant only for scattering at momenta of order $1/\ell$.  
Recall in this connection that we have set the translation parameter $b$ to unity. 
With general $b$, the box excitations come in units of $ b^{-1} 2^{\hat{l}} 2\pi $. So the gap between them, in particular the gap to the first one, $l_\mu=1$, becomes arbitrarily large for sufficiently small $b >0$. But, in the present context, it would not make much physical sense to take $b$ different from order unity. Indeed, the whole idea here is that we cannot probe length scales smaller than of order $\ell$; hence we cannot fix the location of the opaque UV regions, i.e., the boxes, with precision greater than of order $2^{-\hl}$. 

The sum over $\{l_i\}$ is rapidly converging and dominated by the first, $l_{i\mu} = 0$, term, i.e., the range of momenta from zero to the fundamental scale $2\pi/ \ell$ (or $2^{\hat{l}} 2\pi$ in dimensionless units). 
For such `low' momenta (i.e., low compared to $1/\ell$) the vertex (\ref{V3}) becomes indistinguishable from that of the  local $\phi^N$ interaction. 
For momenta of order or larger than  $2^{\hat{l}} 2\pi$, the function $\widehat{\sigma} (k)$ is  very rapidly decreasing.  By (\ref{sfprop1}), (\ref{tens1}),  $\widehat{\sigma} (0) =1$, whereas $\widehat{\sigma} (k)=0$ if any component $k_\mu = 2^{\hat{l}}2\pi l_\mu$ with $l_\mu \not= 0$.  Thus, as seen from (\ref{V3}), 
the vertex kernel is in fact exactly zero if any of the momenta $k_i$ has components exactly tuned to a $2\pi$ multiple of $1/\ell$.

As already note above we must impose the requirement that the FT $\widehat{\sigma}(k)$ of the scaling field  must be entire function in each of its arguments $k_\mu$.  This ensures (perturbative) unitarity provided the operator $K$ in the quadratic part of the action is such as to give ordinary particle propagators (no ghosts) \cite{PS}, \cite{CT1}.  This requirement is automatically satisfied for mother $\sigma(x)$, $\upsilon(x)$  of compact support (e.g., Daubechies wavelets) by the Paley-Wiener theorem.     

It is important to note here that there are no momentum cutoffs present. Integration is over all momenta up to infinity. The interactions though become soft at momenta at and beyond the nonlocality scale $1/\ell$. 
The rapid decay of the vertices (cf. (\ref{DFsfdecay}) for Debauchies wavelets)  ensures UV finiteness for a wide variety of interactions.

As  noted at the end of the previous section, using a tensor wavelet basis results in loss of manifest $SO(d)$ rotational invariance. 
This technical issue, present for $d\geq 4$, will not be further discussed here, since it does not arise in the more general wavelet-inspired framework considered next in section \ref{GF}.

\section{A more general formulation \label{GF}} 
\setcounter{equation}{0}
\setcounter{Roman}{0}

Abstracting from the construction of the model in the previous section 3, 
its essential element  is seen to be an orthogonal projection $\shl{P}$ separating  the coarser from the finer resolution parts of field space relative to a length scale $\ell$. 
Now, given such a $\shl{P}$, the orthogonal compliment can be defined by 
\beq
\shl{Q} \equiv 1 - \shl{P}    \; .   \label{gproj1}
\eeq
Then, $\shl{P}$ being a projection, we have 
\beq
\shl{P}^2 = \shl{P}\,, \qquad \textrm{hence}  \qquad \shl{Q}^2 = \shl{Q} \, , 
\qquad \shl{P} \shl{Q} =\shl{Q} \shl{P} =0  \; .  \label{gproj2}  
\eeq
In the preceding sections we constructed $\shl{P}$ from the scaling field of a MRA via (\ref{dPproj}) and, correspondingly, $\shl{Q}$ via (\ref{dQproj}) from the associated wavelets of the MRA. 
We now consider a generalization where a projection $\shl{P}$ is constructed from functions $\sigma(x)$ smearing over scales $\ell$, but which do not necessarily constitute the  scaling field of a complete MRA.

Let $\sigma(x)$ be a function on $\bbR^d$ 
normalized so that 
\beq
\vev{\sigma, \sigma} = \int d^d x |\sigma(x)|^2 = 1     \; . \label{gnorm1}
\eeq
Let, furthermore, $\sigma(x) = \sigma(|x|)$ so that manifest $SO(d)$ invariance is ensured. The function $\sigma(x)$ is assumed to be well localized around the origin but need not be of compact support. 
It may, for example, be a Gaussian or some other real analytic rapidly decaying function of $|x|$. 
It is required that its FT $\widehat{\sigma}(k)$ be an entire function of $k\in \bb{C}^d$. In fact, one may define $\sigma(x)$ by specifying $\widehat{\sigma}(k)$ so that this property is fulfilled; e.g.,  
\beq
\widehat{\sigma}(k) = \exp (- P(k^2))        \; , \label{gsigmaFT}
\eeq
where $P(k^2)$ is a polynomial with positive highest order term coefficient.    

We again define (cf. (\ref{bas1}):
\beq 
\sigma_{\hl n}(x) \equiv 2^{d\, \hl/2} \sigma (2^{\hl} x - n)    \; , \quad n \in \bbZ^d \; .  \label{gsigma-n}
\eeq
Given the 
set of independent functions $\sigma_{\hl n}(x)$, we may  construct an orthonormal set $\tilde{\sigma}_{\hl n}(x)$ by orthogonalization, specifically, by symmetric (L\"{o}wdin) orthogonalization, which treats all elements of the independent set on an equal footing.\footnote{ The perhaps more familiar  Gram-Schmidt orthogonalization procedure is sequential and not suitable here. The L\"{o}din basis was first introduced in quantum chemistry, where it is widely used.}  It, furthermore, has the property of being, among all orthogonalizations, the closest (in the $L_2$ sense) to the original independent set. 

Let $S_{n, n^\prime}= \vev{\sigma_{\hl \,n}, \sigma_{\hl \,n^\prime}}$ be the overlap matrix. $S$ is hermitian and, as easily shown, a positive definite  matrix.  
Note that, since each $\sigma_{\hl n}(x)$ is well-localized around $2^{-{\hl}} n$, the overlap  between $\sigma_{\hl n}$ and $\sigma_{\hl n^\prime}$ becomes rapidly negligible with increasing $|n- n^\prime|$, so that only a few neighboring $\sigma_{\hl n^\prime}$ contribute to any extent in the matrix elements of $S$. 
The set 
\beq
\tilde{\sigma}_{\hl n}(x) = \sum_{n^\prime \in \bbZ^d} \sigma_{\hl n^\prime} S^{-1/2}_{n^\prime n} \label{L1}
\eeq
then constitute the L\"{o}wdin orthonormal basis; as easily verified, the functions $\tilde{\sigma}_{\hl n}(x)$ satisfy the orthogonality relations (cf. (\ref{orel1})):  
\beq
\int d^dx\;  \overline{\tilde{\sigma}}_{\hl n}(x) \tilde{\sigma}_{\hl k}(x)   = \delta_{nk}  \; .  \label{gorel1}   
\eeq

For the special case of $\sigma(x)$ of compact support contained in the closed ball $B_1 \in \bb{R}^d$ of unit diameter centered at the origin, the functions (\ref{gsigma-n}) 
have support on the ball $B_{\ell} \in \bb{R}^d$ of diameter $\ell$ centered at $x=2^{-\hl} n$. The balls touch but do not overlap, and so we simply have  $\sigma_{\hl n} = \tilde{\sigma}_{\hl n}$. We may in fact more generally replace the hypercubic lattice $\bbZ^d$ by the lattice $\Lambda_d$  achieving closest packing of spheres in $d$ dimensions. Thus $\Lambda_3 \cong A_3 \cong D_3$ is the fcc lattice, 
$\Lambda_4 = D_4$, etc.\footnote{Lattice optimal packings are known for all $d$ up to $d=8$, all by laminated lattices \cite{CS}.} In any case, the formalism in this section accommodates such different choices of $\Lambda_d$.
The Fourier transform of such functions of compact support are necessarily entire functions.   
A prototype example of such functions are the $C^\infty$ `bump' functions: 
\beq
\sigma_{\hl n}(x) =  \left\{ \begin{array}{l l } N \exp \left[\D  -  \frac{\ell^2}{\ell^2- 4|2^{\hl}x - n|^2}\right]  &  |x| \leq 
|2^{-\hl}(n \pm \frac{1}{2}\ell)|    \\
0 &   |x| \geq |2^{-\hl}( n \pm \frac{1}{2}\ell)|    
\end{array} \right.    \; ,  \label{cgsigma-n} 
\eeq
where $N$     
such that (\ref{gnorm1}) is satisfied.   

One should note that the functions $\{\tilde{\sigma}_{\hl n}(x)\}$ defined here do not necessarily constitute a set of scaling functions of a full MRA.  
Since, however, they satisfy the orthogonality relations (\ref{gorel1}), they still define a projection operator in the same manner as 
(\ref{dPproj}), i.e., 
\beq
\shl{P}(x,y)  = \sum_n\tilde{ \sigma}_{\hl n}(x) \overline{\tilde{\sigma}}_{\hl n}(y)      \label{gPproj} 
\eeq 
so that the relations (\ref{gproj2}) are satisfied. The configuration space $\cS$  of a general field $\phi$ is thus decomposed in the orthogonal subspaces $\shl{V}$ and $\shl{\mathcal{W}}$:    
\beq
\vphi = \shl{P} \phi \; , \qquad \chi= \shl{Q}\phi = (1- \shl{P}) \phi  \; . \label{gfexp1} 
\eeq
In particular, (\ref{gfexp1}) implies that $\vphi$ can be written as linear combination of $\tilde{\sigma}_{\hl n}$, and thus, ultimately, of the  $\sigma_{\hl n}$.  
As just noted, the orthogonal compliment subspace $\mathcal{W}_{\hl}$ is no longer assumed to be necessarily spanned by an associated complete wavelet set.\footnote{It is, however, always possible to have a frame decomposition of $\mathcal{W}_{\hl}$.}

We now proceed to define field theories exactly as in sections \ref{UVs}. 
Thus, we introduce interactions for the fields $\vphi$ with action (\ref{S1});  
 include $\chi$ as free fields so as to rewrite the theory in terms of $\phi$; and, proceeding  exactly as in section 3, end up with the same effective Feynman rules. These consist  of ordinary propagators and nonlocal vertices of type (\ref{V3}), which, however,  now arise from the insertion of (\ref{gPproj}) 
in terms of the functions $\tilde{\sigma}_{\hl n}$.
As in the previous section then, interactions become smeared over distance scale $\ell$ and thus insensitive to details at scales $< \ell$. 
With the symmetric orthogonalization $\tilde{\sigma}_{\hl n}$ given by (\ref{L1})  a computation of a $N$-point vertex again yields (\ref{V2})  but now with vertex factor 
\beq
 \lh{V}_N(k_1, \ldots, k_N, l_1, \ldots, l_N) = \prod_{i=1}^N \widehat{\sigma} \big(2^{-\hl} k_i +  2\pi l_i\big) \, \overline{\widehat{\sigma}}(2^{-\hl}k_i)\, \big|t(2^{-\hl} k_i)\big|^2   \; .   \label{V4}
\eeq
In (\ref{V4}) $t(k)$ is given by the trigonometric series 
\beq
t(k) = \sum_{j\in \bbZ^d} c_j e^{ikj}  \; ,    \label{t1} 
\eeq
where the coefficients 
\beq c_j = S_{(n-j)\,n}^{-1/2}          \label{t2}
\eeq
are in fact $n$-independent by virtue of the fact that all $\sigma_{\hl n}(x)$ are obtained from the same function $\sigma(x)$ by translation and by the translation invariance of $\bbZ^d$. The series is rapidly converging with 
$c_0 \sim 1$ and rapidly decreasing $|c_{j \not= 0}|$ due to the small overlaps.

Though still wavelet-inspired, the procedure in this section does away with the need of having the full machinery of a $d$-dimensional MRA. 
The advantage of this is that manifest global $SO(d)$ invariance can be incorporated at the outset. Furthermore, 
we can, in particular, demand  exponential fall-off of interaction vertices in momentum space along the Euclidean direction. This is satisfied, e.g., by (\ref{gsigmaFT}),  (\ref{cgsigma-n}). UV finiteness and perturbative unitarity obtain exactly as before.  
The resulting Feynman rules are of the type also encountered in string field theory.

\section{Concluding discussion \label{DC}} 
\setcounter{equation}{0}
\setcounter{Roman}{0}

In summary, a general scheme was developed for incorporating the physical requirement of loss of resolution inside some  fundamental length scale $\ell$ in a mathematically well-defined and natural way. The formalism is based on the mathematics of wavelets, which  
allows separation of length scales in orthogonal decompositions.
In section \ref{UVs} decompositions within a  complete MRA were employed. The more general formulation of section \ref{GF}, however,  relies only on the construction of a projection operator built from scaling fields that do not necessarily constitute those of a full MRA.

This wavelet-inspired formalism can be applied within any field theory. It results in the nonlocality of interactions implicit in any such loss of resolution within a scale $\ell$. This delocalization of interactions occurs over  regions of extent of order $\ell$. These regions (Euclidean spacetime `atoms'), though not assigned any particular structure, acquire an excitation spectrum in units of $1/\ell$. This is in fact a consequence of the underlying translation invariance of the theory. Interactions inside these regions fall rapidly along the Euclidean directions controlling UV behavior. These features are dictated  by the formalism in a straightforward and natural way.

The resulting delocalization of interactions would generally be expected to lead to some acausal effects. 
Such effects can be estimated to be (exponentially) small outside regions of size of order $\ell$. 
We leave the detailed derivation and discussion  of such estimates to a separate treatment.   
We only remark here that, contrary to being  unwelcome,  such small acausal effects might in fact result in novel interesting features of trans-Planckian physics.

There are some open issues. One is that, as discussed at the end of section 2, within a complete MRA the use of a tensor basis built from $1$-dim bases results in loss of manifest $SO(d)$ invariance. One possible way of dealing with this is the introduction of rotational collective coordinates for the `atoms' (`boxes'), essentially summing over their orientation, which picks a frame. It would be much preferable to start with radial mother functions at the outset, but one lacks such constructions 
for $d\geq 4$. This appears to be a technical issue that may  be resolved in the future as briefly discussed in section 2. The wavelet-inspired generalized formulation of section \ref{GF} evades the problem by going outside the framework of a complete MRA.

Another issue is that of the asymptotic behavior of amplitudes in the  external momenta after their continuation to Minkowski signature. This is a general issue common to all Feynman rules with nonlocal vertices, including those of string field theory. Vertices rapidly decaying along the Euclidean direction may not do so along the Minkowski axis. This is of course related to their being entire functions, which implies that the usual Wick rotation (closing contours at infinity) is generally not possible, cf. \cite{PS}, \cite{CT1}.  The simple Gausssian $\exp (-k^2)$ provides a  standard example. One possibility is that such diverging large external momenta behavior is  canceled in physical amplitudes by adding  the contribution of different interaction vertices. This is what is expected in string field theory, where such Gaussian-type vertices occur; but this apparently has so far not been explicitly verified even in the simplest case of tree level 2-2 scattering, let alone to all orders. Alternatively, for a general field theory, one may impose convergent large external momenta behavior along both Euclidean and Minskowski directions as a requirement on the vertices (e.g., in the specification of $P(k^2)$ in (\ref{gsigmaFT})). This may then be added to the list of physical requirements restricting the possible interactions \cite{T1}. 
These matters are best discussed in the context of the explicit construction of particular models. 
 
Finally, field theories with local symmetries have not been considered in this paper. The inclusion of gauge interactions requires an extension of the present formalism and will be treated elsewhere.

\setcounter{equation}{0}
\appendix
\renewcommand{\theequation}{\mbox{\Alph{section}.\arabic{equation}}} 

\section{Appendix - Multi-resolution analysis and wavelet decompositions } 

In this Appendix we review some basics of wavelet decompositions starting  in $d=1$. 
The standard theory is developed for $L^2(\bbR)$ but it can be extended to other spaces, including classes of distributions 
suitable for  Euclidean path integral fields (cf. below). 
An {\it orthogonal wavelet decomposition}   
is based on a multi-resolution analysis of   $L_2(\mathbb{R})$.

 A {\it multi-resolution analysis} (MRA) of a Hilbert space $\cal{H}$ consists of the following elements: \\
 (i) A sequence of nested subspaces 
\beq 
\cdots V_{-3} \subset V_{-2} \subset V_{-1} \subset V_0 \subset V_1 \subset V_2 \subset V_3 \subset \cdots \label{mR1}
\eeq 
with 
\beq 
\overline{\bigcup_{j\in \mathbb{Z}} V_j}  = \cH  \;, \qquad  \overline{\bigcap_{j\in \mathbb{Z}} V_j}  =\{ 0\} \; .
\label{mR2}
\eeq
(ii) The requirement 
\beq f(x)  \in V_0 \quad \Longleftrightarrow \quad f(2^j x)  \in V_j   ;,     \label{mR3}
\eeq
i.e., that the spaces $V_j$ are scaled versions of one of the spaces, which is conventionally  taken to be $V_0$. 
\newline
(iii) $V_0$ is required to be invariant under integer translations, i.e., 
\beq
f(x) \in V_0   \quad  \Rightarrow \quad f(x-n) \in V_0 \quad \mbox{all} \quad n \in \mathbb{Z}   \; ; 
\label{mR4}
\eeq
(iv) There exists an element ${\rm s} \in V_0$ such that the set 
\beq 
\{ \rs_{0,n} \,| \ n\in \mathbb{Z} \}   \qquad \mbox{is an orthonormal basis in } \ V_0  \; .  \label{mR5}
\eeq
Here the notation 
\beq 
\rs_{jn}(x) = 2^{j/2} \rs(2^j x - n)   \;,   \qquad    \ j,n \in \mathbb{Z}   \label{s_n}
\eeq
is used. 

There are two important immediate implications. (\ref{mR4}) together with (\ref{mR3}) imply that 
\beq
f(x) \in V_j \quad \Rightarrow \quad f(x- 2^{-j} n) \in V_j,\qquad  \mbox{all} \quad n \in \mathbb{Z}  \; ;  \label{mR4a}
\eeq
and (\ref{mR5}) together with  (\ref{mR3}) imply that 
\beq
\{ \rs_{j,n} \,| \ n\in \mathbb{Z} \} \qquad \mbox{is an orthonormal basis in } \ V_j  \quad \mbox{for all} \quad j\in \mathbb{Z}   \; .  \label{mR5a}
\eeq
It is also worth noting  that there may be many subspace ladders that satisfy (\ref{mR1}) - (\ref{mR2}) but do not constitute a MRA without the additional crucial requirements (ii) - (iii).  

For each $j\in \mathbb{Z}$ one now defines the space $W_j$ to be the orthogonal compliment of $V_j$ in $V_{j+1}$: 
\beq
V_{j+1}= V_j \oplus W_j     \; . \label{W1}   
\eeq
Note that it follows that $W_j \perp W_{j^\prime}$ for $j\not= j^\prime$, since, if, e.g., $j > j^\prime$, one has 
$W_{j^\prime} \subset V_j$ and $V_j \perp W_j$. 
Furthermore, as it is easily seen, the spaces $W_j$ inherit the scaling property (\ref{mR3}) of the $V_j$ spaces, so that 
\beq
f \in W_0\Longleftrightarrow   f(2^jx) \in W_j  \;. \label{W2}
\eeq
Suppose now that there is an element $\rw \in W_0$ such that the set $\{ \rw_{0,n} \,| \ n\in \mathbb{Z} \}  $ is an orthonormal  basis for $W_0$, where we defined 
\beq 
\rw_{jn}(x) = 2^{j/2} \rw(2^j x - n)   \;,   \qquad  \ j,n \in \mathbb{Z}   \; . \label{w_jn}
\eeq
(\ref{W2}) then ensures that $\{ \rw_{jn} \,| \, n \in \mathbb{Z}\}$ will be an orthonormal basis in $W_j$. 
The function $\rs(x) = \rs_{00}(x)$ is called the {\it scaling field} and $\rw(x) = \rw_{00}$ is the {\it mother wavelet}. 
Correspondingly, we have the set of scaling fields $\rs_{jn}$ and wavelets $\rw_{jn}$.

Iterating (\ref{W1}) starting from any $j\in \mathbb{Z}$,  (\ref{mR2}) implies  the orthogonal decomposition 
\bal
\cH & = V_j \oplus W_j\oplus W_{j+1} \oplus W_{j+2} \oplus \cdots  \nonumber \\
&= V_j \bigoplus_{j^\prime \geq j} W_{j^\prime}   \;    . \label{W3}
\end{align} 
Similarly,  going the other way,  
\bal 
V_j & = V_{j-1} \oplus W_{j-1} = V_{j-2}  \oplus W_{j-2} \oplus W_{j-1} \nonumber \\
& =  \bigoplus_{j^\prime < j} W_{j^\prime}   \; . \label{W4}
\end{align} 
By (\ref{mR2}) then we have 
\beq
\cH = \bigoplus_{j\in \mathbb{Z}} W_j     \; . \label{W5}
\eeq
Hence, by (\ref{W5}) and (\ref{mR2}) the set $\{ \rw_{jn} \,| j,n \in \mathbb{Z}\}$ is an orthonormal basis for $\cH$. 
Corresponding to these decompositions we have the orthogonality relations:
\bal 
\int \overline{\rs}_{jn} \rs_{jm}  & = \delta_{nm}    \label{orth1}   \\
\int \overline{\rs}_{jn} \rw_{j^\prime m} & =0 \; , \qquad  j^\prime= j +k \, , \quad 0\leq k  \in \bbZ   \label{orth2} \\
\int \overline{\rw}_{jn} \rw_{j^\prime m}  & =\delta_{jj^\prime}  \delta_{nm}    \label{orth3}
\end{align}  
Note that scaling functions $\rs_{jn}$ with different values of $j$ are {\it not} orthonormal. Similarly, wavelets 
$\rw_{mn}$ with $m < j$ are not orthogonal to $\rs_{jn}$ and are not members of the basis in (\ref{W3}). 

By (\ref{W3}) we have the representation
\beq 
f = \sum_{n\in \mathbb{Z}}< \rs_{jn}, f> \rs_{jn}  +  \sum_{\substack{k, n \in \mathbb{Z} \\ k\geq j }}  <\rw_{kn},f> \rw_{kn}   \label{exp1}
\eeq
for any $f\in \cH$.  The decomposition (\ref{W3}) of $\cH$ is an orthogonal decomposition in successively finer resolutions. In the resulting expansion (\ref{exp1}) the basis in $V_j$ allows representation of features of $f$ down to scale $1/2^j$, the $W_j$ basis represents features down to scale $1/2^{j+1}$ that cannot be represented at scale $1/2^j$, the $W_{j+1}$ basis represents features down to scale $1/2^{j+2}$ that cannot be represented on the coarser $1/2^{j+1}$ scale, and so on. 

By (\ref{W4}) - (\ref{W5}) this is equivalent to the alternative expansion 
\beq 
f = \sum_{j, n \in \mathbb{Z} }  <\rw_{jn},f> \rw_{jn}    \;. \label{exp2}
\eeq 
It should be pointed out though that one must be rather careful in using (\ref{exp2}) \cite{D}. This is because it makes sense only in $L^2(\bbR)$ and apparent inconsistencies may arise in various manipulations if, for example, one does not note that it does not converge in the $L^1(\bbR)$ sense. No such `paradoxes' occur with (\ref{exp1}), which holds in most reasonable spaces 
(cf. below). In any event (\ref{exp2}) is practically never used since the need for such decomposition (to infinitely long scales) never arises in physical applications. (\ref{exp1}) is indeed the sole basis of all the discussion in the main text. 

The quantity 
\beq 
P_j(x,y) = \sum_{n\in\bbZ} \rs_{jn}(x) \overline{\rs}_{jn}(y)      \label{Pproj}
\eeq
is the orthogonal projection to the subspace $V_j$. Each subspaces $V_j$ is in fact a self-reproducing Hilbert space with $P_j$ as the self-reproducing kernel. 
Similarly, the orthogonal projection to $W_j$, the orthogonal compliment of $V_j$ in $V_{j+1}$, is given by 
\beq
Q_j(x,y) =\sum_{n\in\bbZ} \rw_{jn}(x) \overline{\rw}_{jn}(y)  \; ;    \label{Qproj}
\eeq
and each $W_j$ is a self-reproducing Hilbert space with (\ref{Qproj}) as its self-reproducing kernel. One then has 
\beq
P_j + Q_j = P_{j+1}  \; . \label{projsum}
\eeq
These properties follow directly from the orthonormality relations (\ref{orth1}) - (\ref{orth3}) and (\ref{W1}), (\ref{exp1}). 

To implement this framework one needs to construct the appropriate scaling field and associated orthogonal mother wavelet satisfying all the above requirements. It is far from obvious that such a MRA actually exists.\footnote{An exception is the Haar wavelet, the simplest example of a wavelet, which had been known since the early 1900's, though, of course, not formulated within the modern MRA framework.} The surprising explicit construction of such orthonormal wavelet bases with good localization properties in the 1980's led to the subsequent explosive development of the subject and its wide range of applications.

If the mother scaling and wavelet defining a MRA are $C^r(\bbR)$ functions with bounded derivatives up to order $r$ and  appropriate decay properties\,\footnote{For compactly supported wavelets such decay properties are of course trivially satisfied.} they satisfy several basic properties. For the scaling field:\footnote{In (\ref{sfprop1} - (\ref{sfprop4}) the constant $c = \int \rs(x) dx$ is normalized to unity for brevity; $c\not=1$ can of course be trivially restored.}   
\bal
\hat{\rs}(2\pi l) &= \delta_{0l}\; , \qquad l\in\bbZ   \label{sfprop1} \\
\sum_{n\in \bbZ} s_{0n}(x) &= \sum_{n\in \bbZ} s(x -n)  =1   \label{sfprop2}  \\
\int P_m(x,y) y^k dy &= x^k\; , \qquad 0\leq k\leq r\, , \quad m\in \bbZ  \; .  \label{sfprop4}
\end{align}
(\ref{sfprop2}) shows that $\rs(x)$ gives a partition of unity.  
For the mother wavelet:
\beq
\int dx x^l \rw(x) = 0 \; , \qquad l=0, 1, \ldots, r \, .  \label{wprop1}
\eeq 
Note that this implies:
\beq 
\hat{\rw}^{(l)}(0) = 0\;, \quad l=0,1, \ldots, r \, . \label{wprop1a}
\eeq
An easy corollary of this, e.g.,  \cite{D},  is that $\rw$ cannot be  $C^\infty$ {\it and} have exponential or faster decay. Thus compact wavelets necessarily have bounded regularity (finite $r$).

Equations (\ref{s_n} and (\ref{w_jn}) are special cases of the general expressions 
\beq 
\rs_{jn}(x) = a^{pj} \rs(a^j x - bn)   \;,   \qquad    \ j,n \in \mathbb{Z}   \label{s_ng}
\eeq
and 
\beq 
\rw_{jn}(x) = a^{pj} \rw(a^j x - bn)   \;,   \qquad  \ j,n \in \mathbb{Z}   \;  \label{w_jng}
\eeq
with general dilation (scale) parameter $a$ and translation parameter $b>0$. In the above the theory was outlined with the standard choices $a=2$ and $b=1$.
The normalization parameter $p$ is irrelevant to the basic theory and usually conveniently chosen to preserve a particular norm under changes of the scale factor; the common choice $p=1/2$, which we adopt here, preserves $L_2$ norms.  
The choice of $a$ for obtaining an orthonormal wavelet bases, on the other hand, is restricted. One has to have rational $a >1$. The standard choice $a=2$ is adopted also here, there being no reason to consider any other value. The translation parameter is not thus restricted. If (\ref{s_n}), (\ref{w_jn}), with scaling and wavelet functions now renamed $\tilde{\rs}$, $\tilde{\rw}$,  
give an orthonormal basis, then so do (\ref{s_ng}), (\ref{w_jng}) with 
\beq
\rs(x) = b^{-1/2} \tilde{\rs}(x/b) \;,    \qquad \rw(x) = b^{-1/2} \tilde{\rw}(x/b)  \; . \label{brel}
\eeq

The above theory for $d=1$ can be extended to multi-dimensional wavelet decompositions. This can be done in three ways \cite{M}, \cite{D}.   
The most immediate way is to take the tensor basis of the $1$-dimensional bases $\rs_{jn}$, $\rw_{j^\prime n}$, $j^\prime\geq j$. This results in $d$ different dilation indices $m$. A better, and in fact the standard way is to take the tensor product of $1$-dimensional MRAs, which gives the tensor product basis (\ref{tens1}) and (\ref{tens2}) 
with one dilation index $m\in \bbZ$. The third and most general way is to define a MRA  directly in $d$-dimensions, i.e., without each space $V_j$ being assumed to be a direct product of lower dimensional spaces \cite{M}. 
This then gives the general $d$-dimensional basis (\ref{bas1}) - (\ref{bas3}).

As already pointed out, wavelet basis decompositions hold in many other spaces than $L^2$ \cite{M}. In particular the above theory gives unconditional bases for all $L^p(\bbR)$, $(1< p <\infty)$.\footnote{As it is well-known, $L^1$ does not admit unconditional bases; so wavelet expansions cannot of course provide one, but still generally do better than Fourier series \cite{M}.} The same holds for the Sobolev spaces $W^s(\bbR)$. For application to 
Euclidean path integral fields extension to classes of tempered distributions is appropriate. 
The space $\cS_r$ consists of all $C^r(\bbR)$ functions $f$ of rapid decay, i.e., 
\beq 
|f^{(k)}| \leq C (1 + |x|)^{-p}   \; , \qquad 0\leq k \leq r\, , \quad \mbox{all} \ p\in \bbZ^+    \label{Sr}
\eeq 
for some $k,p\,$-dependent constant $C$. Compact $C^r$ scaling fields and wavelets are then in $\cS_r$. 
The space of tempered distributions $\cS^\prime_r$ is defined as the dual space (space of continuous linear functionals) to $\cS_r$. It can then be proved that with $\rs \in \cS_r$ there exist a MRA whose union is dense in $\cS^\prime_r$. As a result (\ref{exp1}) holds and converges to $f$ in the sense of $\cS^\prime_r$, i.e., in the distributional sense \cite{W}. It is worth noticing that, in contrast,  the expansion (\ref{exp2}) generally does not necessarily similarly converge to $f$ \cite{W}. 

\end{document}